\documentclass[superscriptaddress,aps,prl,twocolumn,groupedaddress,showpacs]{revtex4}
\usepackage{graphics}
\usepackage{times}
\begin{document}

\title{Novel Quantum Criticality in CeRu$_2$Si$_2$ near Absolute Zero Observed by Thermal Expansion and Magnetostriction}

\author{J. Yoshida}
\author{S. Abe}
\affiliation{Department of Physics, Kanazawa University, Kakuma-machi, Kanazawa 920-1192, Japan}
\author{D. Takahashi}
\affiliation{Department of Physics, Kanazawa University, Kakuma-machi, Kanazawa 920-1192, Japan}
\affiliation{Low Temperature Physics Laboratory, RIKEN, Wako 351-0198, Japan}
\author{Y. Segawa}
\author{Y. Komai}
\author{H. Tsujii}
\author{K. Matsumoto}
\author{H. Suzuki}
\affiliation{Department of Physics, Kanazawa University, Kakuma-machi, Kanazawa 920-1192, Japan}
\author{Y. \={O}nuki}
\affiliation{Graduate School of Science, Osaka University, Toyonaka, Osaka 560-0043, Japan}

\date{\today}

\begin{abstract}
We report linear thermal expansion and magnetostriction measurements for CeRu$_2$Si$_2$ in magnetic fields up to 52.6 mT and at temperatures down to 1 mK.
At high temperatures, this compound showed Landau-Fermi-liquid behavior: The linear thermal expansion coefficient and the magnetostriction coefficient were proportional to the temperature and magnetic field, respectively.
In contrast, a pronounced non-Fermi-liquid effect was found below 50 mK.
The negative contribution of thermal expansion and magnetostriction suggests the existence of an additional quantum critical point.
\end{abstract}


\maketitle
A quantum phase transition (QPT), a continuous order-disorder transition at zero temperature, is driven by quantum fluctuations and differs essentially from a classical phase transition, which occurs due to thermal fluctuations of an order parameter \cite{Sachdev1999}.
A 4{\it f}-based heavy-fermion (HF) system is suitable for studying quantum critical behavior and has attracted experimental and theoretical attention.
The ground state in this system, determined by the competition between the on-site Kondo interaction and the inter-site Ruderman-Kittel-Kasuya-Yosida interaction, can be tuned by physical parameters such as pressure, elemental substitution, and magnetic field.
Furthermore, near a quantum critical point (QCP),
quantum fluctuations cause a pronounced deviation from the Landau-Fermi-liquid (LFL) behavior, leading to what is referred to as non-Fermi-liquid (NFL) behavior. 
In the traditional picture \cite{Hertz1976, Mills1993, Moriya1995}, quasiparticles retain their itinerant character and form a spin-density wave (SDW) type of antiferromagnetic (AFM) order.
This itinerant model has successfully explained experimental critical phenomena close to QCPs \cite{Kambe1997}.
However, recent experiments have shown that this model fails in some HF systems \cite{Schroder2000, Custers2003}.
Thus, a new type of QCP associated with the decomposition of quasiparticles has been proposed \cite{QiSi2001, Coleman2001}.

It has recently been shown that the thermal expansion coefficient $\alpha$ is more singular than the specific heat $C$ in the approach to a QCP. Accordingly, the Gr{\" u}neisen parameter $\Gamma \propto \alpha/C$ diverges as temperature goes to zero near the QCP and the associated critical exponent of $\Gamma$ can be used to determine the type of QCP \cite{Zhu_l}.
In addition, the Gr{\" u}neisen parameter and the thermal expansion coefficient exhibit a characteristic sign change close to the QCP \cite{Garst_e}.
Thus, the Gr{\" u}neisen parameter and the thermal expansion coefficient are widely used as excellent tools for the investigation of QCPs in various systems, for example, HFs \cite{Donath, Kuchler_CeInSn}, spin-ladders \cite{Lorenz}, and perovskite ruthenate \cite{Gegenwart_SrRuO}.

The intermetallic compound CeRu$_2$Si$_2$, which has a ThCr$_2$Si$_2$-type crystal structure, is well known as a typical HF compound, and exhibits the LFL state with an electronic specific heat coefficient $\gamma=$ 350~mJ/K${}^2$mol below the Kondo temperature $T_{\text K}=$ 14~K \cite{Besnus_e, Regnault_e}.
Neutron-scattering measurements represent short-range time-fluctuating AFM correlations described by three incommensurate wave vectors even below $T_{\text K}$ \cite{Rossat_e}.
The elemental substitutions Ce$_{1-x}$La$_x$Ru$_2$Si$_2$ and Ce(Ru$_{1-x}$Rh$_x$)$_2$Si$_2$ induce an incommensurate SDW ground state at a finite transition temperature.
With decreasing $x$, the transition temperature decreases to zero at the finite critical concentrations $x_{c}=$ 0.08 in Ce$_{1-x}$La$_x$Ru$_2$Si$_2$ and $x_{c}=$~0.03 in Ce(Ru$_{1-x}$Rh$_x$)$_2$Si$_2$ \cite{Raymond1997, Tabata1998}.
Since these substitutions induce an expansion of the lattice parameter, CeRu$_2$Si$_2$ is regarded as an example of a pressure-driven QPT with a critical pressure $p_{c}=$ $-$0.3 GPa \cite{Flouqet2005}.
In the elemental substituted system close to the critical concentration,
several critical phenomena down to 0.1 K have been explained by the 3D spin fluctuations of the itinerant model \cite{Tabata1998, Kambe1996}.
In contrast to this traditional picture, our previous experimental results for undoped CeRu$_2$Si$_2$ at small magnetic fields and ultralow temperatures have shown the NFL behavior,
that is, an ac susceptibility peak, and a temperature-independent but field-dependent ac susceptibility and magnetization below the peak temperature of ac susceptibility, $T_P$ \cite{Takahashi_e}.
The critical exponent of the susceptibility below 10 mK could not be explained by the itinerant model, and the suppression of susceptibility by the magnetic field showed a resemblance to the magnetic-field-tuned QCP observed in YbRh$_2$Si$_2$ and CeCu$_{1-x}$Au$_x$ \cite{Schroder2000, Custers2003}. 
In this Letter, we report thermal expansion and magnetostriction measurements for CeRu$_{2}$Si$_{2}$ in magnetic fields up to 52.6 mT at temperatures down to 1 mK.
We present the material's unique critical properties and discuss the existence of an additional QCP.

A single crystal of CeRu$_2$Si$_2$, grown along the {\it c} axis by the Czochralski pulling method using the starting materials Ce (99.99\% pure), Ru (99.99\% pure), and Si (99.999\% pure), was cut into a cylinder, 3~mm in diameter and 5 mm in length.
The sample was cooled with a copper nuclear demagnetization refrigerator and a ${}^3$He-${}^4$He dilution refrigerator.
The thermal expansion and magnetostriction along the {\it c} axis were measured by the capacitance method, with applying the magnetic field in the same direction.
The bottom of the sample was glued to the sample cell made of oxygen-free high-conductivity copper using Arzerite VL-10 silver paste for thermal conduction, and the top was epoxied to the 5-mm-diameter copper capacitance plate using Stycast 2850GT for electrical insulation.
The distance between this plate and another fixed capacitance plate was set to be 18~$\mu$m.
The change in length $\Delta L$ of the sample was derived from the capacitance change.
The sensitivity of $\Delta L/L$ was 10$^{-11}$ by the home-made capacitance bridge that was used.
The thermal relaxation time observed in the experiment was less than several minutes around 10 mK.
The measurements were done by continuously changing temperature and magnetic field 
at a constant rate from $0.4$ to $4$ mK/hour and $5.6$ mT/hour, respectively.
 We observed no hysteresis in temperature and field sweep directions.
Since the thermal expansion and magnetostriction of CeRu$_2$Si$_2$ are 4 orders of magnitude larger than those of Cu below 10 K \cite{Kroeger_e, Fawcett_e}, the change in length of the copper holder was negligible.

%
\begin{figure}[b]
\includegraphics{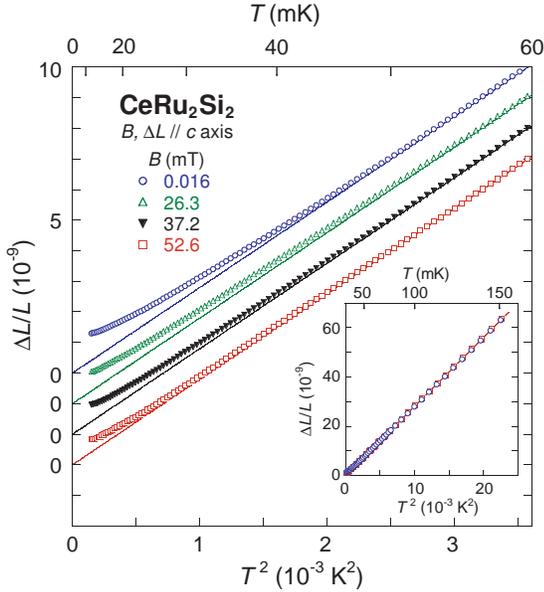}
\caption{(color online). Linear thermal expansion $\Delta L(T)/L$ of CeRu$_2$Si$_2$ along the {\it c} axis at different applied fields, plotted as a function of $T^{2}$. The data are subsequently shifted by 1$\times$10$^{-9}$.The solid lines represent $\Delta L(T)/L=aT^{2}/2$ with $a=5.60\times10^{-6}$~K$^{-2}$.The inset shows the temperature dependence of $\Delta L(T)/L$ at high temperatures.}
\end{figure}
Figure 1 shows the temperature dependence of the linear thermal expansion $\Delta L(T)/L$ in magnetic fields up to 52.6~mT.
As shown in the inset of Fig.\ 1, the linear thermal expansion shows $T^2$ dependence, varying as $\Delta L(T)/L = aT^{2}/2$ in all measured fields above 50~mK.
This temperature dependence indicates that the system is in the LFL state, and the coefficient $a=5.60\times10^{-6}$~K$^{-2}$ obtained agrees with the results of Lacerda {\it et al.}\ \cite{Lacerda2_e}.
However, with decreasing temperature, $\Delta L(T)/L$ deviates from the $T^2$ dependence in all measured fields.
The deviation temperature $T^{*}$ is about 50~mK at 0.016~mT and is suppressed slightly with increasing magnetic field.
In a similar temperature range, the susceptibility shows a Curie-like increase from the temperature-independent Pauli paramagnetic susceptibility \cite{Takahashi_e}.
Thus, we believe that $T^{*}$ is the characteristic temperature of crossover from the LFL to the NFL state, and $\Delta L_{\text{cr}}(T)/L \equiv \Delta L(T)/L-aT^{2}/2$, shown in the inset of Fig.~2, is the critical contribution accompanied by the NFL behavior.
The critical contribution of the linear thermal coefficient $\alpha_{\text{cr}}(T)$ is obtained by the temperature derivative of $\Delta L_{\text{cr}}(T)/L$.
The temperature dependence of $\alpha_{\text{cr}}/T$, shown in the main frame of Fig.~2, differs from $1/\sqrt{T}$ dependence, which is expected from the 3D spin fluctuations based on the itinerant model \cite{Zhu_l}.
In particular, the sign of $\alpha_{\text{cr}}$ remains negative down to the lowest temperature, in contrast to the ordinary pressure-driven QPT of HF systems.

%
\begin{figure}[b]
\includegraphics{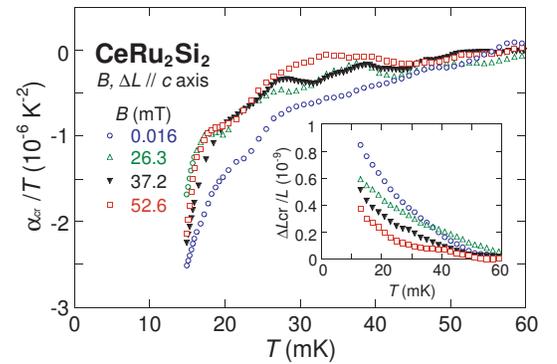}
\caption{(color online). Critical contribution of the linear thermal expansion coefficient $\alpha_{\text{cr}}(T)$ of CeRu$_2$Si$_2$, plotted as $\alpha_{\text{cr}}/T$ vs $T$. The inset shows the critical contribution of the linear thermal expansion $\Delta L_{\text{cr}}(T)/L$ as a function of temperature.}
\end{figure}
In the control parameter vs temperature phase diagram with a symmetry-broken phase extended to a finite temperature from a QCP, the isentropes have a minima on the phase boundary and at the QCP \cite{Garst_e}.
In the case of a pressure-driven QPT, the sign of $\alpha _{\text{cr}}$ is the same as that of the slope of the isentrope and changes on the phase boundary, according to the Maxwell relation $\left( \partial V /\partial T \right) _p =-\left( \partial S /\partial p \right)_T$. 
Most Ce-based NFL materials have an AFM transition line at the low pressure side of the QCP in the $p$-$T$ phase diagram.
Thus, $\alpha_{\text{cr}}$ is positive when the pressure is higher than the phase boundary and negative only inside the AFM ordered state.
In the $p$-$T$ phase diagram of CeRu$_2$Si$_2$, the AFM transition line ends the AFM QCP at $-$0.3 GPa from the substituted results \cite{Flouqet2005}, and no phase transition has been observed at ambient pressures down to 170~$\mu$K~\cite{Takahashi_e}. 
Thus, undoped CeRu$_{2}$Si$_{2}$ at ambient pressure apparently stays on the high pressure side of the AFM QCP, and $\alpha _{\text{cr}}$ should be positive down to the lowest temperature.
One possible explanation for our negative critical contribution $\alpha _{\text{cr}}$ is that an additional QCP exists in the region of higher-than-ambient pressure.
Assuming that the undoped CeRu$_2$Si$_2$ at ambient pressure is located closer to this additional QCP than to the AFM QCP,
the isentropes have a negative slope owing to the isentrope minima of the additional QCP, and thus, $\alpha _{\text{cr}}$ is negative down to the lowest temperature.

%
%
Magnetostriction is a good measure of QCPs, since the magnetostriction coefficient $\Lambda$ is related to the pressure derivative of magnetization via $\Lambda=d[\Delta V(B)/V]/dB=-(\partial M/\partial p)_{T,H}$ on the basis of the Maxwell relation, where $\Delta V(B)/V$ is the volume magnetostriction and $M$ the magnetization per unit volume.
Figure 3 shows the isothermal linear magnetostriction coefficient $\Lambda (B) = L^{-1}d[L(B)-L(0)]/dB$ as a function of the magnetic field. 
Above 50 mK, $\Lambda$ varies linearly with magnetic field, having a linear field coefficient $\lambda=d\Lambda/dB\sim-$9.6$\times$10$^{-7}$ T$^{-2}$.
In magnetic fields higher than those used in our measurements, $\Lambda$ has shown a linear field dependence with a positive coefficient below 6 T and decreased to roughly zero as the field decreased to zero \cite{Lacerda2_e, Paulsen_e}.
The linear field coefficient can be estimated to be $\lambda \sim$ 1.0$ \times$ 10$^{-5}$ T$^{-2}$ in the field range between 2 and 6 T at 70 mK \cite{Matsuhira1999}.
Thus, the linear field coefficient is expected to change from negative to positive in a higher field.
The negative sign with small absolute value of $\lambda$ observed in our low-field measurements implies that magnetization increases with pressure,
which is expected given the quasiparticle mass enhancement in the approach to the QCP.
Thus, the negative magnetostriction is consistent with the existence of an additional QCP at a higher-than-ambient pressure, as suggested by the negative thermal expansion coefficient.

%
\begin{figure}[b]
\includegraphics{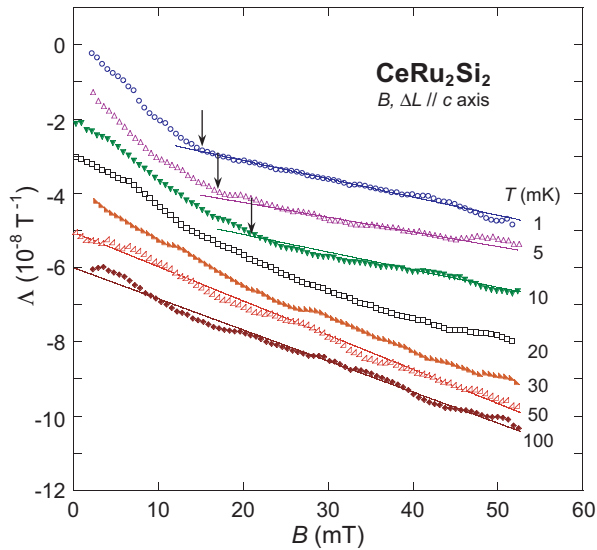}
\caption{(color online). Isothermal linear magnetostriction coefficient $\Lambda(B) = L^{-1} d[L(B)-L(0)]/dB$ of CeRu$_2$Si$_2$ as a function of magnetic field. The solid lines represent the linear field dependence. For clarity, the data are shifted down by 1$\times$10$^{-8}$ T$^{-1}$ at each temperature increment. Arrows show the crossover field to the linear field dependence.}
\end{figure}
The additional QCP indicated by the thermal expansion and the magnetostriction at millikelvin temperatures in CeRu$_{2}$Si$_{2}$ has a different cause than the AFM QCP.
CeCu$_{6}$ also shows a novel behavior at ultralow temperatures. 
Because the AFM transition temperature is suppressed to zero at the critical concentrations $x_{c}=$~0.1 in CeCu$_{6-x}$Au$_{x}$ and $x_{c}=$~0.09 in CeCu$_{6-x}$Ag$_{x}$ \cite{Schlager, Lohneysen2, Heuser}, CeCu$_{6}$ at ambient pressure is located on the high pressure side of the AFM QCP.
In contrast, CeCu$_{6}$ exhibits an AFM order at 2 mK, accompanied by the susceptibility peak and the jump in the thermal expansion coefficient, following the increase of susceptibility from the Pauli paramagnetic susceptibility below 100~mK \cite{Tsujii_e}. 
This behavior has been explained by AFM spin fluctuations.
In the case of CeRu$_{2}$Si$_{2}$, the observed NFL behavior in susceptibility, magnetization, and $\alpha_{\text{cr}}$ below 50~mK can hardly be explained by spin fluctuations based on the itinerant model, and the negative sign of $\alpha_{\text{cr}}$ and $\lambda$ leads to the existence of an additional QCP.

%
We will discuss the possible origin of the additional QCP.
Yamagami and Hasegawa have proposed five Fermi surfaces on the basis of the band structure calculation \cite{Yamagami_e}.
Indeed, the following four Fermi surfaces have been observed experimentally: a large hole surface with 200\,$m_{0}$, a multiconnected electron surface that has crossed arms with 10-20\,$m_{0}$, and two small ellipsoidal hole surfaces with effective mass 1.5\,$m_{0}$ and 1.6\,$m_{0}$, respectiverly, where $m_{0}$ is the free electron mass \cite{Aoki1993, Yano_e}.
The well known HF behavior around 1~K is mainly ascribed to the hybridization between the conduction electrons and $4f$ electrons ($cf$ hybridization) on the large hole Fermi surface because of its large effective mass. We speculate that the effect of the $cf$ hybridization between $4f$ electrons and other Fermi surfaces with a small effective mass 
(``small'' Fermi surface) is not strong enough to contribute to the quasiparticle mass enhancement around 1~K, but this residual hybridization becomes dominant with decreasing temperature.
Owing to the spin fluctuations, the $cf$ hybridization at the small Fermi surface further enhances the quasiparticle mass and the NFL behavior and may induce the additional QCP.

%
The nature of the ground state in a magnetic field has been observed in the magnetostriction at lowest temperatures.
As can be seen in Fig.~3, below 10~mK, a linear field dependence of $\Lambda$, indicative of the LFL state, is observed above 30~mT.
However, $\Lambda$ deviates from the linear field dependence below 30~mT, which suggests the NFL state.
The crossover field between the NFL and LFL states, indicated by the arrow in Fig.~3, decreases as the temperature is reduced.
Such field-tuned crossovers are observed in many HFs.
In the case of CeRu$_2$Si$_2$, the spin fluctuations due to the residual hybridization with the ``small'' Fermi surface that dominated below 50~mK cause $\Lambda$ to deviate from the linear field dependence in the low-field region,
which agrees with the magnetization enhancement observed in similar magnetic field and temperature regions \cite{Takahashi_e}.
In the fields above the crossover field, the magnetic field suppresses the spin fluctuations, which decreases the slope of $\Lambda$.

Finally, we address the critical contribution of the Gr{\" u}neisen parameter, which is an important physical quantity characterizing quantum critical phenomena, as discussed in the introduction. 
The critical contribution of the Gr{\" u}neisen parameter can be expressed as $\Gamma_{\text{cr}}=V_{m}\alpha^{\text{V}}_{\text{cr}}/\kappa C_{\text{cr}}$, where $V_{m}$ is the molar volume, $\alpha^{\text{V}}_{\text{cr}}$ the critical contribution of the volume expansion coefficient, $\kappa$ the isothermal compressibility, $C_{\text{cr}}$ the critical contribution of the specific heat. 
Since the quadrupolar split of Ru nuclei prevents observation of the quasiparticle mass increase in specific heat measurements below 10~mK, we have reanalyzed previous magnetization results \cite{Takahashi_e} to estimate $C_{\text{cr}}$ in a similar way using the Maxwell relation by Paulsen {\it et al.}\ and Sakakibara {\it et~al.}\ \cite{Paulsen_e, Sakakibara_e}.
The static magnetization has shown saturation below a peak temperature $T_{P}$ \cite{Takahashi_e},
and the normalized magnetization can be scaled by the normalized temperature.
On the basis of the temperature dependence of magnetization, as given by $M=M_{s} + \beta(B)(T/T_{P})^2$, below $T \leq 0.6 T_{P}$, 
a negative $\beta(B)$ is obtained at all measured fields and $\beta(B)$ asymptotically approaches zero with increasing field.
The field dependence of $C_{\text{cr}}$ is calculated from the equation 
$\partial(C_{\text{cr}}/T)/\partial B=\partial^{2}M/\partial T^{2}=2\beta/{T_P}^2$.
Thus, the $C_{\text{cr}}/T$ enhancement that corresponds to a decrease in field from 6.21 to 0.20~mT is found to be 26$\pm$14~mJ/K$^2$mol.
The anisotropic volume change has been reported in the LFL state around 1K \cite{Lacerda2_e}. 
However, the temperature region and the physical phenomenon we have discussed are different from the experiment. 
Thus, we assume $\alpha^{\text{V}}_{\text{cr}}=3\alpha _{\text{cr}}$ ignoring the anisotropy at ultralow temperatures. 
The extrapolated value of $\alpha _{\text{cr}}/T$ at zero temperature is $-8~\times~10^{-6}$~K${}^{-2}$ in 0.016~mT.
As a result, $\Gamma_{\text cr}$ turns out to be $-5~\times~10^{3}$.
This growth in the critical contribution of the Gr{\" u}neisen parameter is evidence of quantum critical fluctuation.

In summary,
the ground state of CeRu$_2$Si$_2$ was studied by linear thermal expansion and magnetostriction measurements.
We observed a deviation from the LFL state, indicating that the spin fluctuation is dominant at ultralow temperatures.
The negative critical contribution of the linear thermal expansion and magnetostriction is clear evidence of the existence of an additional QCP.
The dominant spin fluctuation at ultralow temperatures is suppressed by the magnetic field, and a crossover occurs from the NFL state to the LFL state.

%
We thank K.\ Miyake and A.\ Tsuruta for useful discussions, and K.\ Mukai, T.\ Tsunekawa, and K.\ Nunomura for technical assistance. 
This work was supported in part by a Grant-in-Aid for Scientific Research from the Ministry of Education, Culture, Sports, Science and Technology of Japan. 


%
%
%
%

\end{document}